\documentclass[review]{elsarticle}

\usepackage{lineno,hyperref}
\usepackage{graphicx}
\modulolinenumbers[5]

\journal{Journal of \LaTeX\ Templates}

\begin{document}

\begin{frontmatter}

\title{Systematics in the metal-insulator transition temperatures in vanadium oxides}

\author{B. Fisher, J. Genossar and G. M. Reisner}
\address{Physics Department, Technion, Haifa 32000, Israel}




\begin{abstract}
Nine of the known vanadium oxides,  VO$_{2-1/n}$ (n - a positive or negative integer) with n=2 - 6,  8, 9, $\infty$ and -6, undergo metal-insulator transitions accompanied by structural transitions, at various temperatures T$_{MIT}$ (V$_7$O$_{13}$ is metallic above T=0). Among the persistent efforts to determine the driving force(s) of these transitions, electron-electron (Mott-like) and electron-phonon (Peierls-like) interactions, there were several attempts to find systematics in  T$_{MIT}$ as function of n. Here we present an unexpectedly simple and illuminating systematics that holds for positive n:   if T$_{MIT}$ is the absolute value of the difference between T$_M$(n) and T$_P$(n), which represent the contributions of electron-electron and electron-phonon  interactions, respectively, all data points of T$_M$-T$_P$ versus 1/n lie on, or close to,  two simple straight lines;  one is T$_M$-T$_P$= T$_{\infty}$(7/n-1) for V$_3$O$_5$, V$_4$O$_7$, V$_5$O$_9$, V$_7$O$_{13}$, V$_8$O$_{15}$, V$_9$O$_{17}$ and VO$_2$ and the other is T$_M$-T$_P$= T$_{\infty}$(3/n-1) for V$_2$O$_3$, V$_6$O$_{11}$ and VO$_2$.

\end{abstract}

\begin{keyword}

\texttt{A metal - semiconductor}\sep D electron-electron interactions \sep D electron-phonon interactions \sep D phase transitions

\end{keyword}

\end{frontmatter}

\newpage


\section{Introduction}

The general formula VO$_{2-1/n}$ (n - a positive or negative integer) represents all known vanadium oxides. The  monovalent members of the family with n=1 (VO) and n=-2 (V$_2$O$_5$) are metallic and semiconducting at all temperatures, respectively. The monovalent oxides with n=2 (V$_2$O$_3$) and n=$\infty$ (VO$_2$) and the mixed valent oxides  with n=3 - 6, 8, 9 and  -6, undergo metal-insulator transitions (MIT) accompanied by structural transitions, at various temperatures, T$_{MIT}$, below and above room temperature. The mixed valence compounds with positive n belong to the so-called Magneli phases and those with negative n - to Wadsley phases. These transitions, including the absence of a transition for n=7, have attracted attention over many decades. One of the subjects under debate for more than half a century, was the role  of the electron-electron (Mott-like) and electron-phonon (Peierls-like) interactions in  these transitions {\cite{schlog,allred,budai}}. Along with the eforts to understand the nature of these phase transitions there have been several attempts to find a systematic behavior of T$_{MIT}$ as function of n.  In 1971, when only the T$_{MIT}$ of the Magneli phases with n=4 - 8 were known, Nagasawa showed that (1) Magneli phases are classified into
two groups by the transition temperatures in which one has even n and the other odd n and (2) transition temperatures decrease with the increase of even and odd n (the line for the two odd n lies below that connecting the three for even n). In 1980 {\cite{fisher}}, using data from References {\cite{kachi,kartenko}} for n$<$8 we showed that T$_{MIT}$ plotted versus 1/n lie on, or close to several straight lines
as shown below:

(a) T$_{MIT}$=T$_{MIT}$(${\infty}$)(1-1/n) for n= 1, 2, 4 and $\infty$,

(b) T$_{MIT}$=T$_{MIT}$(${\infty})$(1-3/n) for n=5, 6 and $\infty$,

(c) T$_{MIT}$=T$_{MIT}$(${\infty})$(7/n-1) for n=3, 4, 5 and 7.

{\noindent} where T$_{MIT}$($\infty$)=T$_{MIT}$(VO$_2$). Note that the lines (a) and (c) cross at 1/n=1/4 and that the lines (a) and (b) meet at 1/n=0. The then available data hinted at the existence of another simple line, (d)  T$_{MIT}$=T$_{MIT}$(${\infty})$(1-7/n) (the reflection of (c) about the vertical axis 1/n=1/7) for n=7, 8 and $\infty$, but the discrepancy of the data point for n=8 was far too large in view of the close agreement of the points in (a) - (c). At that time, the metallic nature of VO was still problematic, its  non-metallic conductivity  being governed by the large concentration of native defects. In 2007 experimental and computational evidences were presented in support of VO being a strongly correlated metal {\cite{rivadulla}}. The attempt to include the solitary data point with negative n (n=-6) in the systematics  is ignored here.

The renewed inspection of the systematics was motivated by later reports in the literature, including T$_{MIT}$(V$_9$O$_{17}$) {\cite{kuwamoto1,nagata,allred}}  which strengthened confidence in the previously reported systematics.
Two questions were asked at this stage: 1. How valid are these simple lines? 2. If they are valid, what do they imply?

In order to answer the first question, a wide spread of T$_{MIT}$(x) data, where x=2-1/n,  obtained over the years by many groups from electrical and magnetic measurements were tabulated and plotted; these data include thermal hysteresis and experimental errors. Appendix A is devoted to the verification of these lines.

The so proven validity of the data used in References \cite{schlog}, \cite{allred} and \cite{fisher} and the present understanding in the field allowed a  revision of the interpretation of T$_{MIT}$ that simplified the systematics to being expressed by only two straight lines.

\section{Transition temperatures of VO$_x$}

The transition temperatures of the vanadium oxides  shown in Table I were taken mainly from Reference \cite{kachi}; the  data  missing in that early article were filled in from more recent publications. Graphs of T$_{MIT}$(x)  using data from this Table appear in research articles and reviews related to VO$_x$ such as \cite{schlog,perucchi,allred}. A similar graph including  also the datum for V$_9$O$_{17}$ as in \cite{allred},  is shown in Fig. 1  where the primary and secondary horizontal axes have been extended to x=1 and n=1. Solid lines connect the data points as in the above references. The four dashed straight lines which in part overlap with the solid lines represent (a) - (d), their expressions shown  in the previous section.
The datum for V$_9$O$_{17}$ was available in 1981 but was ignored in literature until 2013. As in \cite{fisher}, the point for V$_8$O$_{15}$ deviates significantly from line (d) and that for V$_3$O$_5$ deviates slightly from line (c) (430 K instead of 453 K).

The most interesting pair of lines in Fig. 1 is of course the (c) - (d) pair, for which T$_{MIT}$=$\pm$340(7/n-1)K; these two lines contain most of the data points for T$_{MIT}$ from n=3 to n=$\infty$. Thus,  {\it if the line representing T$_{MIT}$=340(1-7/n)K and the points laying on it (including that for n=8) are rotated around the horizontal axis, all the points of the (c) - (d) pair lie now on a single straight line with the point for n=8 close to it.} This simple transformation leaves out two points on what previously was line (a) (1/n=1/2 and 1) and one point on the former line (b) (n=1/6).  If the point for n=1/6 is reflected about the horizontal axis we obtain a new straight line that connects the points for 1/n=1/2, 1/n=1/6 and 1/n=0.  It turns out that the ordinate of the new graph represents the difference between two functions, while that of the old graph represents the absolute value of this difference. And indeed, the zero for V$_7$O$_{13}$ is not a minimum but a crossover point. The most suitable names for the two functions are T$_M$ and T$_P$ (M for Mott and P for Peierls).
The data points in Fig. 2 represent  T$_M$ - T$_P$ versus 1/n for all V$_n$O$_{2n-1}$ with n$>$1; the solid lines are the fitted linear trendlines with their Patterson correlation coefficients. The dashed lines represent the ideal relations T$_M$ - T$_P$ = 340(7/n -1)K and T$_M$ - T$_P$ = 340(3/n-1)K. The upper dashed line is distinguishable but still very close to the solid line. The lower dashed line is indistinguishable on the scale of  Fig. 2, it overlaps the corresponding solid line. The point for n=1 (VO) remains isolated; the extrapolation of line (a) in Fig. 1  to 0 for n=1 is  not understood.

\section{Discussion}
The discussion on the MIT in V$_n$O$_{2n-1}$ is based on essentials of the state of the art of this topic as described in References \cite{schlog} and {\cite{allred}}. The rutile VO$_2$ structure contains pairs of translationally distinct but symmetry equivalent, parallel, infinite chains of edge sharing octahedra. In V$_n$O$_{2n-1}$ the chains are broken into n-octahedra long units; the symmetry is broken in such a way that rutile-like chain fragments are connected by corundum - like (V$_2$O$_3$-like) chain ends. The two parallel chains become symmetry inequivalent and are denoted in \cite{allred} as "A" and "B". Theoretical calculations for several V$_n$O$_{2n-1}$ compounds \cite{schlog} showed  that electron-lattice interactions are largest in the centers of the chain-fragments while electron-electron interactions are largest at the fragments' ends. Recent experiments \cite{andreev} confirm the importance of electron-electron correlations in V$_3$O$_5$ (the highest point  of Fig. 2). With increasing n the electron-lattice interaction increases while the electron-electron interaction decreases.  The absence  of transition in V$_7$O$_{13}$ is probably due to mutual cancelation.  For V$_2$O$_3$ T$_M$ $>$ T$_P$; this is not surprising since this material was originally considered a canonical Mott system. The location of VO$_2$ at the bottom of Fig. 2 is also not surprising; the structural ordering and pairing at low temperatures imply a dominant Peierls instability.

The initial systematics presented in the Introduction consisted of the three straight lines (a) - (c) and  hinted at line (d) symmetrical to (c). The updated point for V$_9$O$_{17}$ validated line (d) and enabled reduction to two lines to represent the systematics of T$_{MIT}$(1/n). The splitting into two lines occurs between V$_2$O$_3$ and V$_3$O$_5$=V$_2$O$_3$+VO$_2$. T$_{MIT}$(V$_3$O$_5$) and (T$_M$ - T$_P$)(V$_3$O$_5$)) are much higher than those of the constituents. We suggest that this is due to the presence of interfaces between V$_2$O$_3$ and VO$_2$. There are now many examples in the literature showing that the properties of double-, or multi-layers deviate strongly from those of the constituents due to the interface(s). Once the interfaces between V$_2$O$_3$ and VO$_2$ are formed, the addition of rutile fragments allows for a monotonic (linear) variation of (T$_M$ - T$_P$)(1/n).

Rather surprising is the location of the point for V$_6$O$_{11}$ : why should the point for this compound "leave its neighbors" and "join" instead the straight line connecting V$_5$O$_{9}$ to VO$_2$ in Fig. 1 and then fit exactly on the line connecting V$_2$O$_3$ to VO$_2$ in Fig. 2? A hint is provided by the very different dependencies on n of the {\it a} and {\it b} parameters (of chains A and B) as shown in Fig. 6 of Ref. \cite{allred}.  Above n=3, in both the strong and weak electron-electron interaction regimes {\it a} decreases monotonically with increasing x=2-1/n;  the accompanying behavior of {\it b} is at first a weak and then an accelerated decrease  until x=1.83 is reached; above this point {\it b} increases linearly with x towards x=2 (for VO$_2$); a V-shaped {\it a}(x) feature is formed with the tip at n=6 (x=1.83). Electronic structure calculations \cite{schlog} show that electron-electron interactions are stronger in the B chain leading to localization that inhibits electron-lattice coupling. It is possible that the sudden jump of the data point for V$_6$O$_{11}$ out of the upper line in Fig. 2 is correlated with the structural anomaly of {\it b}(x) ; however the exact fit to the lower line is puzzling and requires a stronger argument. If indeed the jump of the data point for V$_3$O$_5$ above that for V$_2$O$_3$ is due to an interface effect, the question now is why  does this effect disappear completely in the case of V$_6$O$_{11}$?

\section{Summary}
The  systematics of T$_{MIT}$ of VO$_{2-1/n}$ as function of n for $1\leq n \leq \infty$ proposed in 1980 \cite{fisher} is revisited here. The renewed  interest was motivated by a  crucial datum that was not available at that time and by the advances in the understanding of the MIT in this family of oxides. These lead to a simplified systematics expressed by two simple expressions;  one is T$_M$-T$_P$= T$_{\infty}$(7/n-1) for V$_3$O$_5$, V$_4$O$_7$, V$_5$O$_9$, V$_7$O$_{13}$, V$_8$O$_{15}$, V$_9$O$_{17}$ and VO$_2$ and the other is T$_M$-T$_P$= T$_{\infty}$(3/n-1) for V$_2$O$_3$, V$_6$O$_{11}$ and VO$_2$ where T$_{MIT}$ is the absolute value of T$_M$-T$_P$. These two expressions emphasize the smooth transition from a Mott-like to a Peierls-like regime of the MIT in this family of materials. The linearity of the difference of the two functions in 1/n, and the existence of two lines, not one, are very intriguing and should be derived from theoretical calculations.

\section{Appendix A}
Most data used in \cite{fisher,schlog,perucchi,allred} are based on Table II (magnetic measurements) of \cite{kachi} while data in Table I of that report (electrical measurements) include thermal hysteresis. Measurements of T$_{MIT}$ of the vanadium oxides have been carried out by many groups. Their data are represented in some cases by a single number, but in most cases by a range of temperatures which include thermal hysteresis or/and experimental error.  In order to answer  question 1. above we collated data for T$_{MIT}$(n) starting from 1936 (see Table II). The limiting temperatures of the transitions either determined by hysteresis or by experimental error are plotted in Fig. 3
,  by open circles for  V$_8$O$_{15}$ and by full circles for all other oxides.  A straight line was fitted to the data points of each of the four groups of oxides (a') - (d'). The resulting  equations and Patterson correlation coefficients (R$^2$) are shown below:

(a') T$_{MIT}$ =337.3(1-1.02/n)K for 1/n = 1, 1/2, 1/4 and 0, R$^2$=0.993

(b') T$_{MIT}$ =339.4(1-2.98/n)K for 1/n = 1/5, 1/6,  and 0, R$^2$=0.997

(c') T$_{MIT}$ =316.5(7.09/n-1)K for 1/n = 1/3,1/4, 1/5 and 1/7, R$^2$=0.998

(d') T$_{MIT}$ =339.3(1-6.97/n)K for 1/n = 1/7, 1/9 and 0, R$^2$=0.999.

Lines (a) - (d) which represent the systematics are  good approximations to the lines (a') - (d') fitted to the collected experimental data of the corresponding groups.

\newpage
\section{Figure Captions}
\begin{figure}[hp]
\includegraphics[scale=0.8]{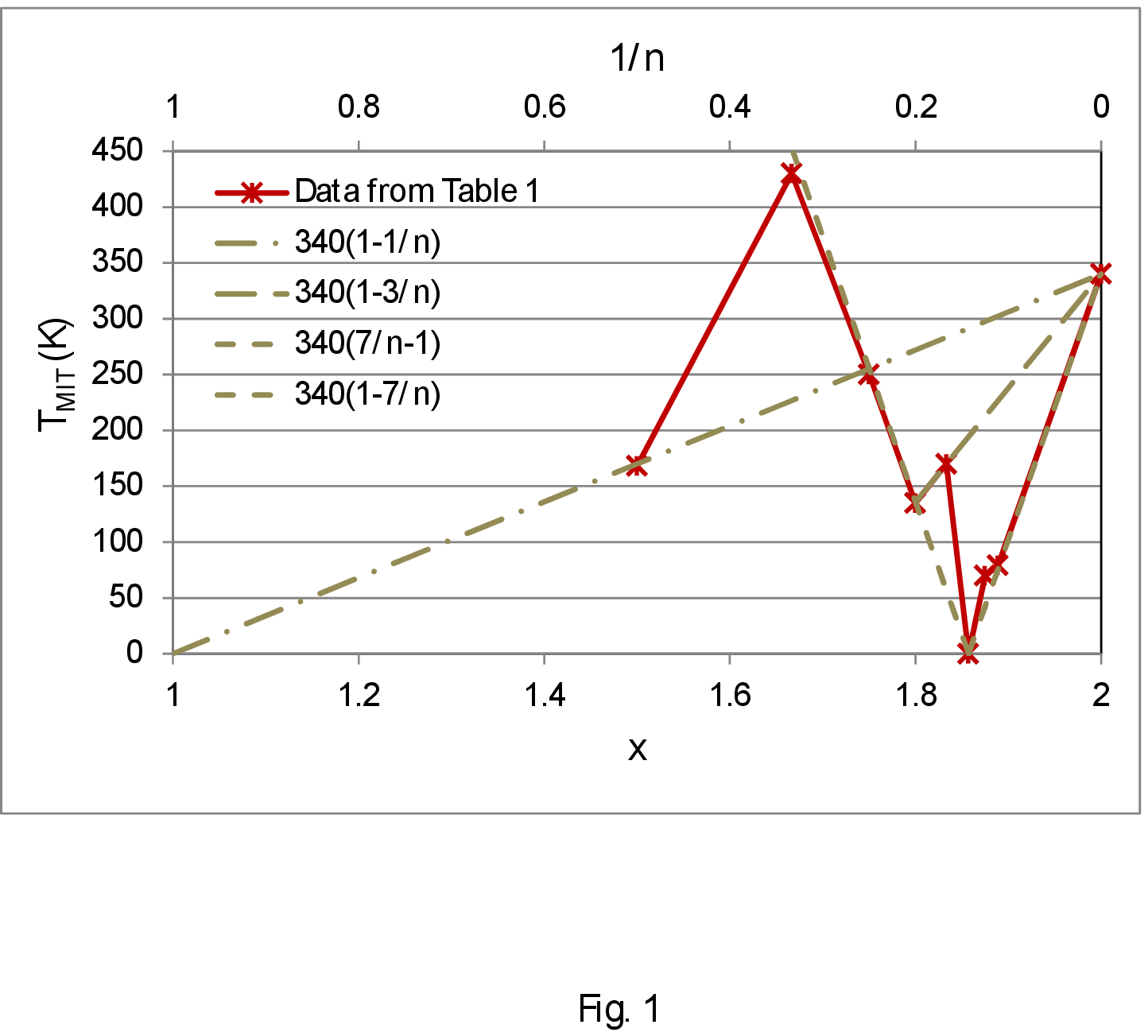}

\caption{\label{1} T$_{MIT}$ as function of x (primary horizontal axis) and 1/n (secondary horizontal axis). Symbols represent data from Table 1 connected by solid lines and dashed lines represent the systematics.    }
\end{figure}

\begin{figure}[hp]
\includegraphics[scale=0.8]{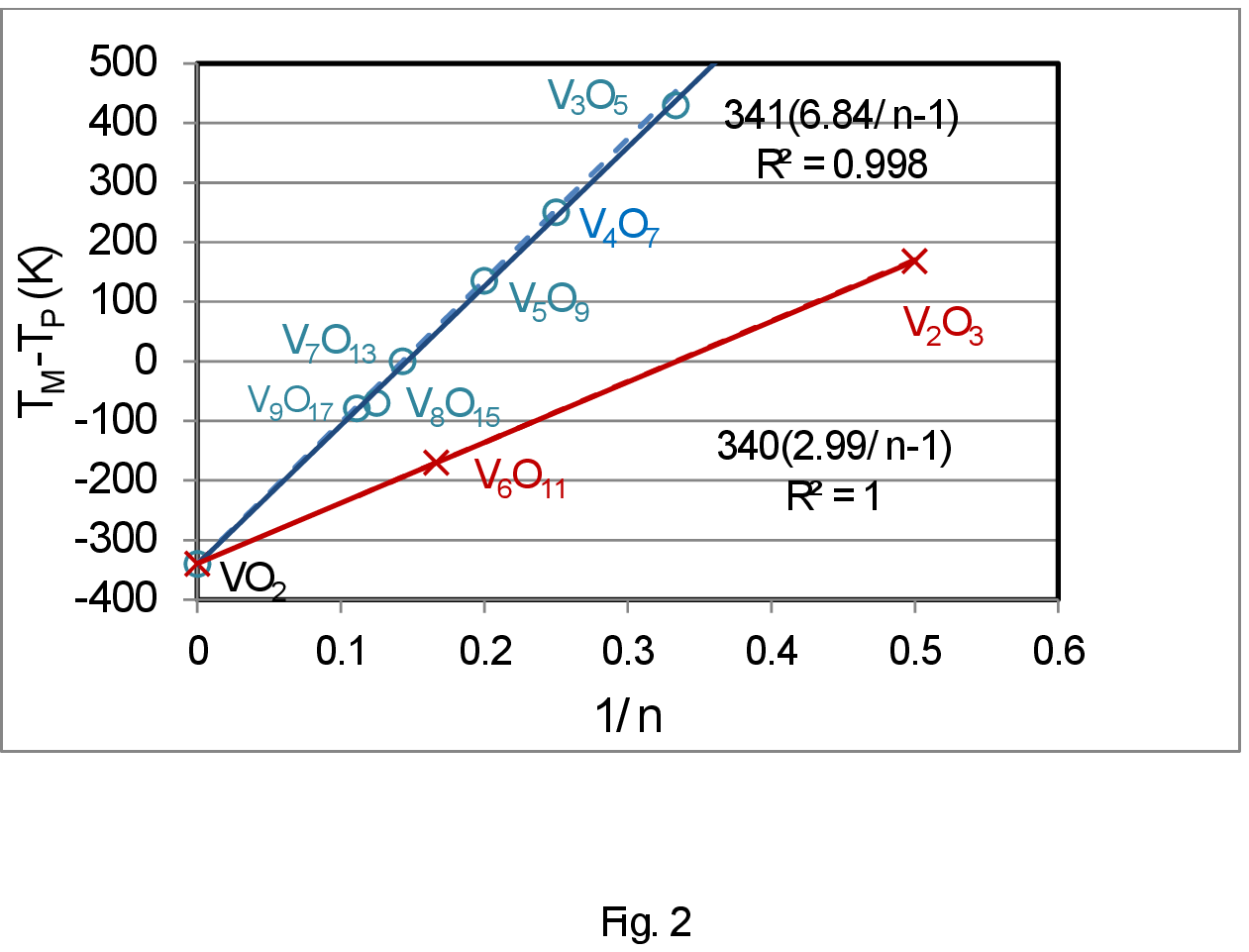}

\caption{\label{2} T$_M$ - T$_P$ as function of 1/n. Symbols represent data points,  solid lines represent linear trendlines fitted to the data and dashed lines represent the two simple expressions of T$_M$ -T$_P$. Note that the lower dashed line is indistinguishable from the corresponding solid line.    }

\end{figure}

\begin{figure}[hp]
\includegraphics[scale=0.8]{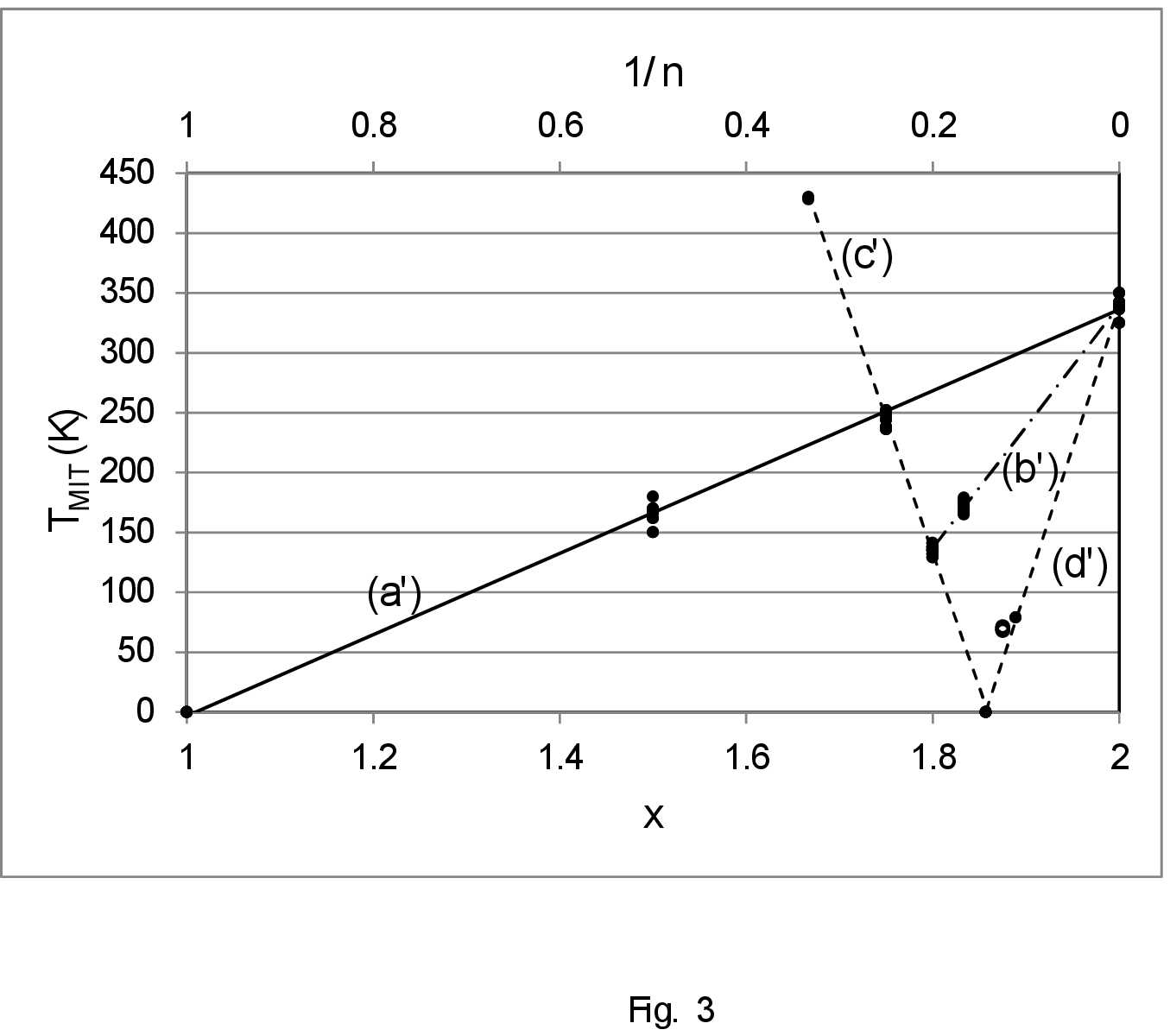}

\caption{\label{3} T$_{MIT}$ as function of x (primary horizontal axis) and 1/n (secondary horizontal axis). Symbols represent data from Table 2, straight lines are fitted linear trendlines  to the (a') - (d') groups of data. }

\end{figure}



%


\begin{table}
 \caption{\label{1} Vanadium oxides V$_n$O$_{2n-1}$ (positive n) and their M-I transition temperatures (T$_c$)}
\begin{tabular}{cccc}

 n&V$_n$O$_{2n-1}$&T$_{MIT}$ (K)&Reference\\
 \hline \hline
 1&VO&metallic&{\cite{rivadulla}}\\
 \hline
 2&V$_2$O$_3$& 168& {Table II in \cite{kachi}}\\
 \hline
 3&V$_3$O$_5$& 430& {\cite{kartenko}}\\
\hline
4&V$_4$O$_7$& 250& {Table II in \cite{kachi}}\\
 \hline
5&V$_5$O$_9$& 135& {Table II in \cite{kachi}}\\
\hline
6&V$_6$O$_{11}$& 170& {Table II in \cite{kachi}}\\
\hline
7&V$_7$O$_{13}$& metallic&  \cite{kachi}\\
\hline
8&V$_8$O$_{15}$& 70& {Table II in \cite{kachi}}\\
\hline
9&V$_9$O$_{17}$& 79&\cite{nagata} \\
\hline
$\infty$&VO$_2$& 340& {Table II in \cite{kachi}}\\
\hline
\end{tabular}
\end{table}

\newpage
\begin{table}
 \caption{\label{2} Vanadium oxides V$_n$O$_{2n-1}$ (positive n) and their M-I transition temperatures (T$_c$)}
\begin{tabular}{cccc}

 n&V$_n$O$_{2n-1}$&T$_{MIT}$ (K)&Reference\\
 \hline \hline
 1&VO&metallic&{\cite{rivadulla}}\\
 \hline \hline
2&V$_2$O$_3$& 170&\cite{anderson}\\
\hline
2&V$_2$O$_3$& 162 - 180& \cite{foex}\\
\hline
 2&V$_2$O$_3$& 153 - 165& {\cite{Morin}}\\
 \hline
  2&V$_2$O$_3$& 168$\pm 2$&{\cite {kosuge}}\\
 \hline
 2&V$_2$O$_3$&150 - 162   &{\cite{mcwhan}}\\
 \hline
 2&V$_2$O$_3$& 168& {\cite{kachi}}\\
 \hline
 2&V$_2$O$_3$&150 - 170 & {\cite{kuwamoto}}\\
 \hline\hline
 3&V$_3$O$_5$& 430&{\cite{chudnovski}}\\
 \hline
3&V$_3$O$_5$& 428& {\cite{andreev}}\\
\hline\hline
4& V$_4$O$_7$& 250$\pm 2$& {\cite{kosuge,nagasawa}}\\
\hline
4& V$_4$O$_7$&250, 244 - 250 & {\cite{nagasawa,kachi,honig}}\\
\hline
4& V$_4$O$_7$& 237$\pm 1$ & {\cite{andkli}}\\
\hline\hline
5&V$_5$O$_9$&139$\pm 2$&\cite{kosuge}\\
\hline
5&V$_5$O$_9$&135 $\pm 3$& \cite{nagasawa}\\
\hline
5&V$_5$O$_9$&135, 129 - 135& \cite{kachi,honig}\\
\hline\hline
6&V$_6$O$_{11}$& 177$\pm 2$& \cite{kosuge}\\
\hline
6&V$_6$O$_{11}$& 170$\pm 3$& \cite{nagasawa}\\
\hline
6&V$_6$O$_{11}$&170, 174 - 177& \cite{nagasawa,kachi,honig}\\
\hline
6&V$_6$O$_{11}$& 165 - 170&\cite{andkli1}\\
\hline\hline
7&V$_7$$O_{13}$& metallic& \cite{kosuge,nagasawa,kachi}\\
\hline\hline
8&V$_8$O$_{15}$&70, 70$\pm 1$& \cite {nagasawa,kachi,honig}\\
\hline
8&V$_8$O$_{15}$& 69$\pm 1$& \cite {ueda}\\
\hline\hline
9&V$_9$O$_{17}$& 79& \cite{kuwamoto1,nagata}\\
\hline \hline
$\infty$& VO$_2$ & 325 - 350& \cite{Morin}\\
\hline
$\infty$& VO$_2$ & 338.5$\pm 2$& \cite{sasaki}\\
\hline
$\infty$& VO$_2$ & 340$\pm 2$& \cite{kosuge,bongers}\\
\hline
$\infty$& VO$_2$ & 339$\pm 1$& \cite{paul}\\
\hline
$\infty$& VO$_2$ & 340$\pm 1$& \cite{macchesney}\\
\hline
$\infty$& VO$_2$ & 340& \cite{kachi}\\
\hline

\end{tabular}

\end{table}









\end{document}